\documentclass[12pt]{article}
\usepackage{graphicx}
\usepackage{amssymb,amsmath,amsfonts,palatino,amsthm}
\usepackage{amssymb}
\usepackage{epstopdf}
\usepackage{slashed} 
\newcommand{\f}{\begin{equation}}
\newcommand{\ff}{\end{equation}}

\begin{document}

\title{What are we missing in our search for quantum gravity? \\}
\author{Lee Smolin\thanks{lsmolin@perimeterinstitute.ca} 
\\
\\
Perimeter Institute for Theoretical Physics,\\
31 Caroline Street North, Waterloo, Ontario N2J 2Y5, Canada}
\date{\today}
\maketitle

\begin{abstract}
 
Some reflections are presented on the state of the search for a quantum theory of gravity.  I discuss diverse regimes of possible quantum gravitational phenomenon, some well explored, some novel.

{\it For Foundations of Mathematics and Physics : an IJGMMP special issue}

\end{abstract}

\newpage

\tableofcontents


\section{Introduction}

Despite enormous effort from thousands of dedicated researchers over a century\footnote{The idea that there might be quanta of gravitational waves was first mentioned by Einstein in a paper in 1917\cite{AE1917}.}, the search for the quantum theory of gravity has not yet  arrived at a satisfactory conclusion.  
We have indeed several impressive proposals, each of which partly succeeds in describing plausible quantum gravitational physics.  Each tells a compelling story that has, for good reason, won it advocates.  Each has also run into persistent roadblocks, which are pointed to by their skeptics.  Looking back, before strings and loops, before causal sets, causal dynamical triangulations, asymptotic safety, amplitudes, twisters, shape dynamics, etc,  to the early days of Bergman, Deser, DeWitt, Wheeler and their friends, who would have thought that there would turn out to be at least half dozen ways to get part way to quantum gravity?

Perhaps we might, for a moment, consider that the approaches so far pursued are not really theories, in the sense quantum mechanics, general relativity and Newtonian mechanics are theories.  For those are based on principles and perhaps we can agree that we don't yet know the principles of quantum gravity.  There are of course proposals for quantum gravity principles, and part of the reason for this paper is to prepare the  ground for proposals 
of new principles (or, in some cases, such as the holographic principle, sharpening up their formulation)\footnote{Hence, this is a companion paper to \cite{4principles}.}. Instead, let us, just for a moment, think of the current approaches as models, which each describe some plausible quantum gravitational phenomena.  The different approaches can then be thought of as complementary, rather than in conflict, as they investigate diverse regimes of 
possible new physics. Could we hope that taking this view may open up discussions between people working on different approaches, to the benefit of all of us?

This frees us up to consider that, despite genuine achievements on several sides, we have yet to see a real theory of quantum gravity.  Can we then begin to look for one?
If we adopt this view, we can learn from all that has been done, while taking a clean slate.  How then do we proceed to look for a theory?

Of course, we work under an obvious handicap, which is that there are few experiments whose results can guide us by winnowing down the possibilities.  But there are a few real Planck scale experiments, which have yielded clues, and which are on the threshold of constraining possible quantum gravity effects at order of $\frac{\mbox{Energies}}{E_{Pl}}$.  Even now we can be fairly sure that Lorentz invariance is not simply broken at that order\cite{QGphen-review}\footnote{But see \cite{GAC-neutrinos}.}.  There is also a window into possible quantum gravitational effects in cosmology, such as low $l$ anomalies\cite{lowl} or parity breaking in $B$ modes\cite{parityB}.  These represent opportunities that must be explored.  

In situations like this, it can be good to pause and take stock of where we are\cite{Whatis}.  The following are some reflections on what we may be missing in our search for quantum gravity.  

\section{What is missing from attempts to discover quantum gravity?}

Approaches are great, and we have good reasons for affection for particular ones.  But can we put aside the different approaches and, especially, their unfortunate sociologies, and just talk physics?

We can start with a simple question:

\subsection{Where are the zeroth order quantum gravity phenomena?}

In many prior revolutionary  transitions there was a key first step where some well studied phenomena, which are already understood in the then current theoretical framework, were reinterpreted in terms of new concepts and principles.  This often leads to surprising new insights,  by giving us a Rosetta stone for translating between the old and new theoretical languages.  That is, at first the correspondence is a mere reinterpretation of phenomena, already explained by the old theory, in a surprising new language.   

Associated with this dual description is a new parameter, which controls novel phenomena.  The correspondence establishes at order zero in the new parameter a translation  between the languages of the new and old theory.  
But as soon as this is established we notice that the correspondence holds in a limited domain.  There is then a space to move beyond the zone of correspondence to novel phenomena whose scale is set by the new parameter.  By doing so we adventure into a new regime of phenomena, but one with clear connections to established knowledge.

Here are some examples of how a transition to a new theory was initiated by reinterpreting a familiar, well understood phenomena in new terms. 

\begin{itemize}

\item{} Galileo's reinterpretion of the tower experiment\cite{Feyerabend}.  Consider the fact that a ball dropped from the top of a tower falls to the bottom.  This simple fact has two interpretations, which lead to opposite conclusions.  If you are an Aristotelean, you interpret this result as evidence that the earth doesn't move.  But if you are Galileo, and believe in the principle of inertia, you interpret the same result as a confirmation that the Earth could be moving without our experiencing any effects.  After all, he argued, a ball dropped from the top of a ship's mast, while sailing smoothly in the harbour of Venice, falls to the bottom of the mast.

\item{}In special relativity; mass is reinterpreted as energy.  One new phenomena this allows is pair creation i.e. the transformation between matter-energy and other forms of energy.  The new parameter is $\frac{v^2}{c^2}$.  

\item{} In general relativity,  the equivalence principle explains in a radically new way the old fact that all massive objects fall with the same acceleration.  Newton understood this as a consequence of the equality of inertial and gravitational mass, which seemed to be a coincidence. Einstein explained this as a necessary consequence of a new principle. This allows  gravity to be reinterpreted as the curvature of spacetime.  The new parameter is $\frac{GM/c^2}{r}$.

\item{}Matter was initially modelled as a continuum, i.e. fluids, gases and solids.  Boltzmann, Kelvin and others reinterpreted continuous, thermodynamic phenomena in terms of the atomic hypothesis.  At first they were able to work out many correspondences, such as the ideal gas law and the laws of thermodynamics.  These correspondences were exact in the thermodynamic limit, in which Avogadro's number would go to infinity.  Then, Einstein and others noted that if Avogadro's number were finite there would be novel phenomena such as Brownian motion.


\end{itemize}

Let us then ask, {\it Where are the zeroth order quantum gravity phenomena?}   
Can we find zeroth order correspondences between some well known phenomena and quantum gravity?  We don't have a real start on quantum gravity unless we can provide an answer to this.  Here are some proposals for zeroth order quantum gravity phenomena. 

\begin{enumerate}

\item{}{\it  The zeroth order phenomena is locality itself. } This must be the  case if as is sometimes hypothesized, locality is emergent in the classical or continuum limit of a fundamental quantum theory of gravity, whose states are networks living in no space, perhaps spin networks or records of entanglement. The first order departures from locality are quantum phenomena, especially entanglement.   Indeed one version of this proposal is that spatial relations are emergent from entanglement\cite{entanglementgeo,CHM,holegraphic,QMfromQG}.

The second order departures from locality are then disordered locality\cite{disloc} and relative locality\cite{RL1,RL2}.  

\item{} {\it More precisely, the zeroth order quantum gravity phenomena is space itself,} specifically its low dimensionality and fantastically low curvature, compared to the Planck scale.  Julian Barbour used to emphasize that space itself, and especially the low dimensionality is a highly nonlocal phenomena\cite{JB-personal}.  This is seen if you try to express the physics of $N$ particles in terms of their relative distances, $r_{jk}$, alone.  For these $\frac{N (N-1)}{2}$ quantities are determined in terms of $3N-6$ coordinates, which is many fewer.  This means the $r_{jk}$  are subject to a large number, $C= \frac{N(N -7)}{2} +6 $ constraints.  These can be understood as the vanishing of the volumes of all independent $n$-simplices, with $n>3$, made from the $r_{jk}$.

Indeed, the $AdS/CFT$ correspondence succeeds in generating one dimension of space from $d$ others,
in the special case that the cosmological constant is negative
\cite{ADSCFT,CHM,entanglementgeo,holegraphic}.  This provides many interesting examples to study, and provides a partial Rosetta stone for translating between conformal field theory phenomena and gravitational phenomena.  It however remains to be seen whether the construction helps us do what we would really like to do which is to understand how all he dimensions of space emerge from something more fundamental.


\item{}{\it The zeroth order phenomena is gravity.}  This is suggested by the thermodynamic derivations of general relativity by Jacobson\cite{Ted95,Ted2015} and elaborations on it\cite{paddy,ls-spinfoams,4principles}.  In the course of the derivation one refers to quantum phenomena such as the Unruh temperature\cite{Unruh}, so $\hbar$ appears.  Another $\hbar$ appears explicitly in expressing an entropy proportional to area.  These $\hbar$'s cancel in the resulting Einstein's equation.  This applies even more to Verlinde's entropic derivation of Newtonian gravity, in which $\hbar$'s and $c$'s  are both present, but cancel\cite{EVentropic}.

\item{}{\it Could another zeroth order phenomena be MOND?}  Perhaps MOND\cite{MOND} is a quantum gravity effect, for positive cosmological constant, $\Lambda$, which arises in a regime, or phase, of accelerations small compared to 
\f
a_\Lambda = c^2 \sqrt{\Lambda}
\ff
which also involves the cancellation of $\hbar$'s and $c$'s.  
This idea is explored  from diverse perspectives in \cite{4principles} and in \cite{EVdark}-\cite{thermalMOND}.

\item{}{\it The zeroth order phenomena is the universe itself, its vast scale and stability as well as the relative stability of the laws.}

\end{enumerate}

The hypotheses just mentioned may serve as bridges to the true theory of quantum gravity.  While we are looking for such bridges, let's keep in mind theories and hypotheses that are clearly transitional and incomplete, but
nonetheless may capture some of the truth. 

\begin{enumerate}

\item{} The proposal that space is emergent from 
entanglement\cite{entanglementgeo,CHM,holegraphic,QMfromQG}.

\item{} The causal set hypothesis that spacetime is really a discrete causal set made up of discrete events and their causal relations\cite{CS,CS2,ECS1,ECS2,ECS3,Cohl}.  This is very like the hypothesis that matter is made of large but finite collections of atoms.   The first order phenomena would be analogous to Brownian motion.  Two have been proposed: 1) the hypothesis that the cosmological constant is the result of a fluctuation\cite{Sorkin-cc} and 2) covariant dispersion\cite{covdis}.

\item{} The shape dynamics hypothesis that the universe is not a four dimensional spacetime 
mod spacetime diffeomorphisms\cite{SD1,SD2}.  
It is instead reinterpreted as an evolving three dimensional geometry mod spatial diffeorphisms and Weyl transformations (i.e. local conformal rescaling.)  This is analogous to Galileo's reinterpretation of the tower experiment from an Aristotelean demonstration of the Earth's stationarity to a demonstration of the principles of relativity and inertia.

A proposed first order phenomena where the two are no longer different interpretations, but differ substantially is in black hole singularities, which are eliminated in favour of bounces to baby universes in shape dynamics\cite{SD-BH}.

\item{} In relative locality\cite{RL1,RL2} and its discrete version, energetic causal sets\cite{ECS1,ECS2,ECS3}, 
a picture of particle dynamics in which relativistic particles propagate on a fixed background spacetime is replaced by an apparently equivalent picture in which particles propagate on a fixed momentum space.  Interactions which happen locally at spacetime events in the old picture become events, elements of a causal set, at which energy-momentum conservation is imposed.  In this new picture spacetime emerges as an auxiliary description.   The first order phenomena where they diverge is gotten by curving or adding 
torsion or non-metricity to momentum space, which leads to the novel phenomena of relative locality.  

\item{}The $AdS/CFT$ hypothesis in the planar limit in which $N\rightarrow \infty$, is a precise dictionary for translating some classical gravitational phenomena into an equivalent, non-gravitational language\cite{ADSCFT}.  There  are many interesting correspondences.  And there are clear paths for going beyond zeroth order to give novel quantum gravitational phenomena.

It is then urgent to understand this correspondence in terms that both don't rely on a negative cosmological constant and apply very generally, without the need for supersymmetry or special dimensions. Some suggestions to explore are in 
\cite{SDAdS} and also in the companion paper\cite{4principles}, where I suggest that the   $AdS/CFT$ correspondence is an expression of a more general quantum equivalence principle.


\end{enumerate}

\subsection{Phenomenological limits and regimes  of quantum gravity}

Whatever the quantum theory of gravity is, it will depend on four dimensional constants, $\hbar$, $G$, $c$
and $\Lambda$.  We are familiar with the commonsense idea that the limit of $\hbar \rightarrow 0$ with the 
others fixed should define general relativity, while the limit in which $G$ and $\Lambda$ are taken to vanish should give quantum field theory.
But, there are several other interesting limits of the three parameters $\hbar$, $G$ and $c$, which each define a regime of quantum gravity phenomenology.

Two interesting regimes come from taking $\hbar \rightarrow 0$;  while $c$ is held fixed;  these may be called the non-quantum regimes of quantum gravity.  

\subsubsection{The relative locality regime}

We can recall first the relativity locality regime in which $G$ and $\hbar$ are both taken to zero, with $c$ held fixed, but with the Planck mass also held fixed, giving us\cite{RL1,RL2}
\f
m_{p}^2 = \frac{\hbar c}{G}
\ff
This defines a regime of phenomena in which $m_p$ and $c$ are fixed while both quantum and gravitational effects are negligible, because $G=\hbar = 0$.  Since $l_p = \sqrt{\frac{\hbar G}{c^3}}   \rightarrow 0$ there is no quantum geometry.   In this regime the propagation and scattering of particles may be deformed due to curving momentum space\cite{RL1,RL2}.  But there is no corresponding deformation of wave propagation.  Indeed, as $\hbar =0$ the correspondence between waves and particles is lost.

Notice that in this regime the entropy of black holes goes to infinity, while their temperature remains finite\cite{RLbh}.

\subsubsection{The strong gravity regime}

Alternatively we can explore phenomenology where $G \rightarrow \infty$ as $\hbar \rightarrow 0$, such that 
\f
\frac{\hbar G}{c^3} \rightarrow l_p^2, \ \ \mbox{fixed}
\ff
again with $c$ held fixed.  It follows that $m_p \rightarrow 0$.  Now there is no deformation of particle dynamics, while wave propagation can be modified, for example as,
\f
\left ( \frac{2}{c^2} \frac{\partial^2}{\partial t^2}   - \nabla^2 -l_p^2 \nabla^4 + \ldots  \right )\phi =0
\ff

Now the black hole entropy stays finite while the temperature of black holes goes to zero.

An unusual feature of this limit is that it is the opposite of the semiclassical limit.  In this limit the commutation relations of quantum gravity are unchanged, because they involve $l_p^2$.
\f
[ A_b^j (x) ,  \tilde{E}^a_i (y) ] =  \hbar G  \delta_b^a \delta_i^j \delta^3 (x ,y)
\ff

Meanwhile, the commutators of matter degrees of freedom go to zero
\f
[ \phi (x) , \pi (y) ] = \imath \hbar \delta^3 (x ,y) \rightarrow 0
\ff

\subsection{The holographic regime}

A very interesting regime about  which a lot is known is the one studied in most calculations in the $AdS/CFT$ correspondence\cite{ADSCFT}.
This is based on a limit in which one takes $N$ large, where $N$ measures the ratio of the cosmological constant scale to the Planck scale.
\f
N = \frac{R^2}{l_{pl}^2} =\frac{1}{\hbar G \Lambda}
\ff
Here $\Lambda = -\frac{1}{R^2}$.  $N$ can be seen as counting the number of degrees of freedom defined on a boundary in
an asymptotically anti-deSitter spacetime. 

\subsection{The loop quantum cosmology regime}

A second cosmological regime is related to quantum cosmology\cite{LQC,LQC-review}, and is defined by the limit in which the Planck energy
$E_p$ is taken to infinity, while the Planck energy density
\f
\rho_{p}=\frac{E_p}{l_p^3}= \frac{c^7}{\hbar G^2}
\label{rhop}
\ff
is held fixed.The speed of light, $c$ is also kept fixed.  In terms of $\hbar$ and $G$ this limit is defined by scaling by a 
dimensionless $t$
\f
\hbar =\hbar_0 t^4 , \ \ \ \ \ \ \  G= G_0\frac{1}{t^2}
\ff
where $t$ is then taken to infinity, so that $E_p$ and $l_p$ both diverge.

This is justified by the FRW equation or Hamiltonian constraint, which can be read as
\f
( a^\prime )^2 = \frac{\rho^{matter}}{\rho_p}+ \frac{1}{E_p} \int d^3x  (\partial h_{ij} )^2 .
\ff
Here the spatial metric is expanded as 
\f
g_{ij}= a^2 ( \delta_{ij}+ \sqrt{\frac{l_p}{m_p}} h_{ij} )
\ff
and $\prime $ denotes differentiation with respect to dimensionless conformal time.
In the regime defined by holding $\rho_p$ fixed while $E_p$ is taken to infinity the spatial inhomogeneities decouple, 
$g_{ij} \rightarrow a^2  \delta_{ij}$, 
and we are left with the homogeneous equation studied to good effect in papers on loop quantum 
cosmology\cite{LQC,LQC-review}.

\subsection{Are there any Newtonian quantum gravity phenomena?}

The regimes we mentioned previous are relativistic in that the speed of light, $c$, is held fixed.  But, 
are there phenomena which are measured in units of hG or h/G with no $c$'s?  i.e. is there a Newtonian regime of quantum gravity?

Here is a curious fact: combinations of just $\hbar$ and $G$ without $c$ are not simple.  To get simple quantities like a mass or a length we need to combine them with $c$.  Indeed the Planck mass and Planck length involve all three constants, $\hbar$, $G$ and $c$.  This is a simple truism but it means that any of the characteristic phenomena  we associate with $l_p$ and $m_p$, such as the discreteness of quantum geometry or the unification of the forces, will go away in the limit $c \rightarrow \infty$ and so they will not have Newtonian analogues.

So it is worth asking whether there are any characteristic Newtonian quantum gravity phenomena, which occur at scales parametrized by combinations of $\hbar$ and $G$ alone, without $c$.

Indeed, there are such characteristic phenomena associated with combinations of the other two pairs.  
$\hbar $ and $c$ go together well to convert length to momentum or time to energy.  Together with $e^2$ they give us the dimensionless fine structure constant, which organizes the scales of phenomena in quantum electrodynamics.
$G$ and $c$ together convert a mass to a length $R_{Schw}= \frac{2G}{c^2} M $.  
But simple combinations of  $\hbar$ and $G$ don't seem to make anything that parameterizes a new, unexpected phenomena.  

By dimensional analysis, the phenomena exhibited by a Newtonian quantum gravity regime would involve some peculiar physical dimensions.
Without $c$, there would be no Planck mass,  nor is there a Planck length.  There is a 
unit of 
{\it mass per square root of velocity.}
\f
{\cal A}^2  = \frac{\hbar}{G}  = \frac{mass^2}{speed}
\ff
There is also a unit of {\it  length to the fifth per time cubed.}
\f
{\cal B}  = \hbar G = \frac{length^5}{time^3}
\ff
This suggests Lifshitz scaling at small velocities.  Perhaps a connection to MOND\cite{MOND}?

It would be very interesting to discover a regime of phenomena where $c$ has been taken to infinity, but where 
the new quantities $\cal A$ and $\cal B$ are measurable.  Presumably this involves heavy,  slow quantum gravity objects.

Note that it can't involve an analogue black holes as $c$ has gone to infinity.

If we look at more complex combinations of $\hbar$ and $G$, we find a conversion from $mass^{-3}$ to length, given by
\f
r_{g}= \frac{\hbar^2}{G} \frac{1}{m^3}
\ff
This is the ``gravitational Bohr radius", i.e. from the Schroedinger equation, the ground state of an atom held together by a Newtonian gravitational potential has a wave function
\f
\psi (r) = e^{-\frac{r}{r_{g}}}
\ff
A Newtonian gravitational atom would be a good trick to play with, but it wouldn't teach us anything about quantum gravity.  In any case they will be prohibitively big for atoms and small for planets, as 
\f
r_{g}=  l_p \frac{m_p^3 }{m^3}
\ff
with a correspondingly tiny binding energy:
\f
E_g = -\frac{1}{2} \frac{G^2 m^5}{\hbar^2} = - \frac{mc^2}{2} \left ( \frac{m}{m_p} \right )^4
\ff






\subsection{Newtonian quantum cosmology}  

Of course there is another dimensional constant in quantum gravity, the cosmological constant, $\Lambda$ and, with it 
together with $\hbar$ and $G$ one can form a complete set of units without $c$.  These depend on whether you take the fixed constant to be the inverse length-squared $\Lambda=\frac{3}{R^2}$ or the Hubble time $T=H^{-1}$.

There is a third possibility, which is to hold the cosmological
acceleration $a_\Lambda = \frac{c^2}{R}$ fixed.  These three regimes are inequivalent as $c$ has been taken to $\infty$.

Note that we take care to distinguish the empirically measured $MOND$ acceleration,
$a_0$, which is determined to be roughly $1.2 \times 10^{-8} \ cm/s^2$ from the 
cosmological acceleration $a_\Lambda$, which is related to the cosmological constant.

\begin{itemize}

\item{}{\bf Newtonian quantum cosmology: }$\Lambda =\frac{1}{R^2}$ fixed as $c\rightarrow \infty$.
\begin{eqnarray}
m_g &=& \left (  \frac{\hbar^2 }{GR }  \right )^{\frac{1}{3}} = m_p  \left (  \frac{l_p}{R}  \right )^{\frac{1}{3}} =10^{-20} m_p
\\
t_g&=&  \left (  \frac{R^5  }{\hbar G  }  \right )^{\frac{1}{3}} = \frac{R}{c} \left (  \frac{R  }{l_p }  \right )^{\frac{2}{3}}
\end{eqnarray}
where in the right hand expressions the $c$'s cancel.  

Note that $m_p$ is roughly the proton mass!.  This is a $c \rightarrow \infty$ residue of Dirac's large number phenomena.

\item{}{\bf Hubble-time Newtonian quantum cosmology} $T = H^{-1} $ fixed as $c\rightarrow \infty$.
\begin{eqnarray}
r_H&=&  \left (  \hbar G T^3 \right )^{\frac{1}{5}} = R \left (  \frac{l_p  }{R }  \right )^{\frac{2}{5}} = 10^{-24} R
\\
m_H &=& \left (  \frac{\hbar r_H }{GT }  \right )^{\frac{1}{2}} = 10^{-12} m_p  =10^{-20} m_p
\end{eqnarray}

\item{}{\bf MOND quantum cosmology} $a_\Lambda = \frac{c^2}{R} $ fixed as $c\rightarrow \infty$ and $R \rightarrow \infty$.

Here there is a unit of velocity
\f
v_0 = \left ( \hbar Ga_\Lambda^2  \right )^{\frac{1}{7}} = c  \left (  \frac{l_p}{R }  \right )^{\frac{2}{7}} \approx 10^{-7} cm/s 
\ff

We can study a MOND bound atom, which has a potential
\f
U_{MOND} = -\frac{GmM}{r} + m v^2_{TF} \ln {r / \rho_0 }
\ff
where $v^2_{TF}=\sqrt{GM a_0 }$, expressing the Tully-Fischer relation\cite{TF}.  This gives a MOND-Bohr radius of 
\f
r_{MOND} = \frac{\hbar}{m v_{TF} }
\ff
The binding energy is order
\f
E_0 \approx - m v_{TF}^2
\ff
which we note has no $\hbar$ in it.  

\item{}{\bf The classical MOND  limit} $a_0 \approx a_\Lambda= \frac{c^2}{R} $ fixed as $c\rightarrow \infty$, 
$R \rightarrow \infty$ and $\hbar \rightarrow 0$, with $G$ fixed.

The key constant here is
${\cal A}_0 = G a_0$, which is a  conversion between mass and velocity to the fourth power.   This constant is actually measured, by observations of the Tully-Fischer relation\cite{TF}
\f
v^4 = G a_0 M_{b}
\ff
where $M_b$ is the baryonic mass of a galaxy, and $v$ is the stellar rotational velocity in the outer disk where the rotation curve flattens out.
Fits to data find
\f
a_0= 1.2 \times 10^{-10}  ms^{-2}
\ff
which is not far from $a_\Lambda = c^2 \sqrt{\frac{\Lambda}{3}}$. 

Note that this is a limit in which $\Lambda \rightarrow 0$ from above.

A reason we might expect to see novel phenomenon in this limit is that for small accelerations,
\f
a < a_\Lambda
\ff
the equivalence principle need not apply.  One way to say this is that an 
observer's acceleration horizon-the horizon created by their own acceleration, falls near their cosmological horizon when $a < a_\Lambda$.  In \cite{4principles} and \cite{thermalMOND} 
I argue that this could be the origin of MOND.

\end{itemize}

\subsection{Energy and its positivity}

A key issue for any non-perturbative approach to quantum gravity is the role of energy.  The puzzle originates from the equivalence principle, which forbids there from being a local measure of the gravitational field.   This is so that a freely falling observer has no way to distinguish their situation from an inertial observer in Minkowski spacetime, to leading order in separations over curvatures.

As a consequence, the energy of the gravitational field can only be measured quasi-locally, or at infinity.  So there is no expression of the energy of a spacetime, or region thereof, expressed as a volume integral over a positive definite expression.

It turns out that the energy  of the gravitational field is still positive\cite{posADM}. This is very fortunate, otherwise flat empty spacetime would be unstable.  But this positivity is an on-shell property.  It only holds in the presence of the field equations or, in the Hamiltonian formulation-of the constraints.

We the must ask whether there is in quantum gravity an operator on the space of physical states that represents the energy which is both positive definite and Hermitian, in the physical inner product.  Such an operator cannot be just the sum of squares of local operators.

In \cite{posE} I have explored conditions on the physical inner product that must be satisfied if there is to be a positive definite and Hermitian operator representing the $ADM$ mass.

\subsection{Where does the Planck mass come from?}

There is another issue regarding energy that challenges quantum gravity theories.  This is that the Planck area,
$l_p^2 = \frac{\hbar G}{c^3}$ turns up easily and naturally, while the Planck mass 
\f
m_p =\frac{\hbar}{l_p}  .  
\label{mp}
\ff
 does not easily turn up.  The reason
is the following.

The  action principle for the gravitational field in general relativity, including the boundary term, is proportional to $\frac{c^3}{G}$.
\f
S^{gr} =  -\frac{c^3}{8 \pi G} \left [   \int_{\cal M} d^4 x  ( R - 2 \Lambda )  -  \int_{\partial {\cal M}} d^3 \sigma \kappa  
\right ] -   \int_{\cal M} d^4 x  {\cal L}^{matter}
\ff
This is indeed the only place that $G$ appears in the action.  A little dimensional analysis tells us why.  The Riemann curvature scalar, $R$, and the intrinsic curvature $\kappa$ are both purely geometrical.  $R$ has units of inverse length-squared while the intrinsic curvature $\kappa$ has units of  inverse length.  If the action is to be, well, an action, this has to be turned into a mass.  The conversion factor $\frac{1}{G}$ is needed to convert a length into a mass.  The same is true for the boundary term, it has a  $\frac{1}{G}$ in front of it.

The phase factor of the path integral then is of the form 
\f
e^{\frac{\imath S}{\hbar}} =
e^{-  \imath \frac{1}{8 \pi \hbar G} [ \int_{\cal M} d^4 x      ( R - 2 \Lambda ) -  \int_{\partial {\cal M}} d^3 \sigma \kappa     ] 
- \frac{1}{\hbar}  \int_{\cal M} d^4 x  {\cal L}^{matter} } 
\ff
so we see that in the absence of matter, $G$ only appears in the combination $l_p^2 = \hbar G$.  
Without matter, the gravitational action is invariant under a scaling 
\f
\hbar \rightarrow \lambda \hbar, \ \ \ \ G \rightarrow \frac{G}{\lambda}
\ff

The same is true of the commutation relations between the Ashtekar connection, $A_a^i$ and the
inverse sensitized from field, $\tilde{E}^a_j$ which represents information about the three geometry
\f
[ A_a^i (x) ,  \tilde{E}^b_j (y)  ] = -  \hbar G \delta^3 (x,y) \delta_a^b \delta^i_j
\ff
It then seems impossible without matter  to produce the expression (\ref{mp}) for $m_p$.
It is the same for the spin foam action, which differs from a topological field theory by the imposition
of the simplicity constraint.  The latter is a constraint on representations and is dimensionless.

We see the same story when the boundary term comes from the boost Hamiltonian
of  FGP \cite{Ale-BH,Bianchi-BH,ls-spinfoams}, which has dimensions of action (since it is conjugate to a dimensionless boost), and is equal to\footnote{Note that this is the contribution to the action from a corner of a causal diamond, and hence comes into an action directly, without being integrated over time.},
\f
H_B = \frac{c^3}{8 \pi G} A(W)
\ff
where $A(W)$ is the area of the horizon as seen by the boosted observer.

Indeed, in $LQG$ the Planck area and Planck volume appear easily and give the scale of the spectrum of quantum geometry.   But it appears that to talk about the Planck energy, we need to couple the 
quantum geometry to  matter.  Matter terms in the action will bring in independent factors of
$\hbar$..

One place the  Planck mass appears is the energy for the Schwarzchild black hole in the isolated horizon approach\cite{isolated}, which is taken to be
\f
H= M = \frac{c^2}{4\pi G}\sqrt{A} \approx \sqrt{\frac{\hbar}{G}} n \frac{c^2}{4\pi}
\ff
Here $n$ stands for the area quantum numbers.   But this is written down by definition, it is not derived from a Hamiltonian operator defined on the whole Hilbert space.

If we want $LQG$ to make predictions for quantum gravity phenomenology, it has to be able to speak to us about corrections of the form of energies in units of $\frac{1}{m_p} \approx \sqrt{\frac{G}{\hbar}}$.  

This is connected to the assumption that the quantum theory of gravity predicts phenomena associated with gravitons such as the gravitational analogue of the photo-electric effect, at least for wavelengths long compared to $l_p$.   This must ultimately be due to a 
normalization of the linearized hamiltonian which gives to each graviton an energy $\hbar \omega$, which is independent of $G$.   Linearized and
perturbative quantum gravity introduces the Planck mass, when it gives
the perturbed metric, $h_{ab}$, defined by 
\f
g_{ab} = \eta_{ab}+  \sqrt{\frac{G}{c^2}} h_{ab}
\ff
 canonical dimensions of square root of energy per length.  This and the assignment of $\hbar$ to loops separates the dependence on $G$ and $\hbar$. 
This separation occurs also in  the semiclassical approximation \cite{mesemiclass}.
But  I know of no mechanism for producing those factors  from the non-perturbative quantum theory without invoking matter.  

But if $m_p$ is missing in the bulk dynamics of $LQG$ and spin foam models then it may be because these 
describe, not full quantum gravity, but a strong coupling limit of the theory in which $\hbar \rightarrow 0$ and $G \rightarrow \infty$ with $l_p$ held fixed
and $m_p$ taken to zero.  Indeed we can see this, from the form of the Ashtekar connection\cite{Abhay},
\f
A= \Gamma (e) + \imath G \Pi
\ff
so in the limit $G\rightarrow \infty$ we pick up the ultra local limit in which all spatial derivatives go away.

In the light of these comments we can consider the derivations of black hole 
thermodynamics by Perez et al\cite{Ale-BH} and Bianchi\cite{Bianchi-BH}.  At two crucial steps in their derivations 
they introduce $\hbar$ independently when they relate the simplicity constraint to the first law of 
thermodynamics\footnote{See \cite{ls-spinfoams} for discussion of how the first law of thermodynamics plays a role 
in these arguments.}.  This requires identifying $B^a = \hbar K^a$ as the boost Hamiltonian in the Hilbert space of a triangle.  Similarly they identify $T_U=\frac{\hbar}{2\pi c}$ as the boost temperature.   These independent introductions of $\hbar$ make it possible to extract Newton's constant from the ratio $\frac{l_p^2}{\hbar} = G$.  Without this they couldn't derive the classical Einstein equations (with matter) from the quantum statistical physics of the horizon.

\subsection{Gravity is missing}

If $LQG$ and other approaches fail to talk about energy, they fail too when it comes to gravity.  That is, in classical general relativity there is a straightforward way to derive Newton's gravitational theory as the non-relativistic approximation to general relativity.  If one pulls a scalar field, $\phi$ out of the metric by a rescaling 
$g_{ab} \rightarrow g_{ab}^\prime = e^\phi g_{ab}$ it follows right away that 
\f
\nabla^2 \phi = 4 \pi G \rho
\ff

I know of only one way to get this out of the Hamiltonian approach to $LQG$, which is by an indirect entropic 
argument\cite{ls-entropy}.

This is of course consistent with the hypothesis that the Hamiltonian approach to $LQG$ describes quantum general relativity only in the 
strong coupling limit in which $G\rightarrow \infty$ while $\hbar \rightarrow 0$.

The situation is better in spin foam models where one can recover the $\frac{1}{q^2}$ behaviour of the graviton propagator from correlation functions of boundary excitations\cite{graviton}. 

\subsection{Maximal $CPT$ violation}

There is another line of thought that should be connected to this one.  This is the set of arguments that lead to the conclusion that irreversibility and time reversal invariance breaking are fundamental.  These are discussed in \cite{SURT,TR,ECS1,ECS2,ECS3,TA1,TA2}.
These lead to the conclusion that the familiar time symmetric laws hold in a limited regime, beyond which we should see the effects of a preferred fundamental arrow time.  Models for how time reversal physics might emerge in a limited regime from a more fundamental time irreversible physics are described in \cite{ECS1,DDDS1}.

One consideration along these lines begins by noting that according to the $CPT$ theorem,  $CPT$ must be a symmetry of any Lorentz invariant relativistic $QFT$.  But global Lorentz invariance is an accidental or emergent symmetry of the ground state of the gravitational field-Minkowski spacetime.  Thus we may hypothesize that $CPT$ is enforced only to the extent that the assumptions of the $CPT$  theorem hold.  This $CPT$ regime should then be delimited by, $R$ the radius of curvature of spacetime.  Thus we should expect to find $CPT$ violation on the order of 
\f
\Delta^{XCPT}= \frac{\lambda}{R}
\ff
where $\lambda$ is a wavelength.

Here is one idea:  assume the fundamental theory is irreversible, but there is an emergent theory which is a local, lorentz invariant QFT.
Then by the CPT theorem the emergent theory has CPT symmetry.  This suggests that CPT is maximally broken, given that CPT is enforced by the lorentz invariance of the emergent theory.  Now lorentz invariance is broken if the metric is curved, so the CPT breaking should be proportional to the curvature tensor.  So they could be given by effective actions like:
\f
\Delta S \approx \frac{1}{m_p} R_{ab}\bar{\psi} \gamma^a \gamma^b \psi
\ff

The second idea is that my precedence theory of quantum dynamics, introduced in \cite{precedence}, has no need to be time reversal invariant, so if its true we should see rare processes which break time reversal.

\section{A new strategy: quantum gravity as a principles theory}

In light of these reflections we might consider novel strategies for searching for quantum gravity.  One is to stop asking for a specific model of quantum spacetime but, instead, to search for general principles which might constrain the choice of models to investigate.  That is, following Einstein, we seek
a {\it principle theory}, rather then a  {\it constitutive theory.}
This strategy is explored in a companion paper\cite{4principles}.

\section*{ACKNOWLEDGEMENTS}

I am grateful to Joseph Kouneiher for including me in this volume. 
I would like to thank Andrzej Banburski, Linqing Chen,  Bianca Dittrich, Laurent Freidel, Henriques Gomes, 	Jerzy Kowalski-Glikman, Joao Magueijo and Yigit Yargic for very helpful discussions and encouragement.

I am also indebted to
Stacy McGaugh, Mordehai Milgrom and Maurice van Putten for very helpful correspondence.

This research was supported in part by Perimeter Institute for Theoretical Physics. Research at Perimeter Institute is supported by the Government of Canada through Industry Canada and by the Province of Ontario through the Ministry of Research and Innovation. This research was also partly supported by grants from NSERC and FQXi.  I am especially thankful to the John Templeton Foundation for their generous support of this project.

 \end{document}